\documentclass[prd,amssymb,amsmath,amsfonts,aps,nofootinbib,notitlepage]{revtex4-1}
\usepackage{hyperref}
\usepackage{graphicx}
\usepackage{color}

\def\lsim{\raise0.3ex\hbox{$\;<$\kern-0.75em\raise-1.1ex\hbox{$\sim\;$}}}
\def\gsim{\raise0.3ex\hbox{$\;>$\kern-0.75em\raise-1.1ex\hbox{$\sim\;$}}}

\newcommand{\AddrAHEP}{%
  AHEP Group, Institut de F\'{\i}sica Corpuscular --
  C.S.I.C./Universitat de Val{\`e}ncia \\
  Edificio Institutos de Paterna, Apt 22085, E--46071 Valencia, Spain}
\newcommand{\AddrUdeA}{%
   Instituto de F\'{\i}sica, Universidad de Antioquia,
    A.A. 1226, Medell\'{\i}n, Colombia}
\newcommand{\AddrEIA}{%
   Escuela de Ingenier\'{\i}a de Antioquia,
     A.A. 7516, Medell\'{\i}n, Colombia}

\begin{document}
\title{Gravitino dark matter and neutrino masses with bilinear R-parity
  violation} \author{Diego~Restrepo}\email{restrepo@udea.edu.co}
\affiliation{\AddrUdeA} \author{Marco~Taoso\footnote{Multidark
    fellow}}\email{taoso@ific.uv.es}
\author{J.~W.~F.~Valle}\email{valle@ific.uv.es}
\affiliation{\AddrAHEP}
\author{Oscar~Zapata}\email{pfozapata@eia.edu.co}
\affiliation{\AddrEIA}
%
%
\begin{abstract}

  Bilinear R-parity violation provides an attractive origin for
  neutrino masses and mixings.
  In such schemes the gravitino is a viable decaying dark matter
  particle whose R-parity violating decays lead to monochromatic
  photons with rates accessible to astrophysical observations.  We
  determine the parameter region allowed by gamma-ray line searches,
  dark matter relic abundance and neutrino oscillation data, obtaining
  a limit on the gravitino mass $m_{\tilde G} \lsim$ 1-10~GeV
  corresponding to a relatively low reheat temperature $T_R \lsim$ few
  $\times 10^7-10^8$~GeV.
  Neutrino mass and mixing parameters may be reconstructed at
  accelerator experiments like the Large Hadron Collider.

\end{abstract}
\maketitle

\section{Introduction}
\label{sec:intro}

The origin of neutrino masses and mixing and the nature of dark matter
are two of the most elusive open problems of modern particle physics
and cosmology, which clearly indicate the need for new physics beyond
the Standard Model.
It has been suggested that these two apparently unrelated issues may
be closely
inter-linked~\cite{berezinsky:1993fm,lattanzi:2007ux,bazzocchi:2008fh,Esteves:2010sh}.
Here we propose an alternative way to relate dark matter with neutrino
properties within a scenario where supersymmetry (SUSY) is the origin of
neutrino mass~\cite{hirsch:2004he}, thanks to the spontaneous
violation of R-parity~\cite{masiero:1990uj}. For definiteness and
simplicity we adopt an effective description in terms of explicit
bilinear R-parity violating superpotential terms (BRpV)
\cite{Diaz:1997xc,Hirsch:2000ef,diaz:2003as}.

We show how both dark matter and neutrino oscillations can be
simultaneously explained in the presence of bilinear R-parity
violation with gravitino lightest supersymmetric particle (LSP), in such a way that,
\begin{itemize}
\item gravitino dark matter properties are closely related to the
  scale of neutrino mass, and
\item 
 neutrino oscillation parameters may be reconstructed at
  accelerator experiments.
\end{itemize}
Indeed, in this model the very same lepton number violating
superpotential terms that generate neutrino masses and mixing also
induce dark matter gravitino decays, as this also breaks R parity.

We show how, although unprotected by R-parity, the gravitino can be
stable over cosmological times and be a viable cold Dark Matter (DM)
candidate. This follows from the double suppression of its decay rate,
which depends on the small R-parity violating couplings and it is
suppressed by the Planck
scale~\cite{Takayama:2000uz,Buchmuller:2007ui}~\footnote{The gravitino
  as a Warm Dark Matter candidate in the BRpV model with gauge
  mediation and its collider implications have been studied
  in~\cite{Hirsch:2005ag}.  }.
Interestingly, gravitino decays produce monochromatic photons, opening
therefore the possibility to test this scenario with astrophysical
searches.  Requiring the model parameters to correctly account for
observed neutrino oscillation parameters~\cite{Schwetz:2011qt} implies
that expected rates for gamma-ray lines produced by gravitino decays
of mass above a few GeV would be in conflict with the Fermi-LAT
satellite observation~\cite{Abdo:2010nc,Vertongen:2011mu}, leading to
an upper bound on the gravitino DM mass.
The bound on the gravitino mass with bilinear R-parity violating
couplings holds under the assumption of universality in gaugino masses
(see~\cite{Vertongen:2011mu} and references therein). Here we study
the conditions of non-universality in gaugino masses to relax the the
gravitino mass constraints. We show how the bound remains even if we
assume non-universal gaugino masses, though somewhat less stringent.
Turning to the implications at collider experiments such as the Large Hadron Collider (LHC)
these have been discussed in a series of earlier
papers~\cite{Mukhopadhyaya:1998xj,Choi:1999tq,Porod:2000hv,deCampos:2005ri,deCampos:2007bn,DeCampos:2010yu}.
It is important to stress the expected signatures are basically the
same already studied in the usual BRpV scenarios.

In the next Section we briefly introduce the BRpV model, in
section~\ref{sec:gravitino} we discuss gravitino decays and
cosmological relic abundance and explain our numerical procedure.  
In particular we obtain an upper limit for the gravitino mass and
discuss it both for universal and non-universal gaugino masses.
In Section~\ref{sec:nnlsp} we briefly discuss collider implications,
and finally summarize the paper in Sec.\ref{sec:Summary}.

\section{Bilinear R-parity violating model}
\label{sec:BRpVM}

Here we work in a constrained minimal supergravity model in which the
gravitino is the lightest supersymmetric particle (LSP). The simplest
R-parity violation scenario is assumed, in which the superpotential
contains bilinear R-parity violating
terms~\cite{Diaz:1997xc,Hirsch:2000ef}
\begin{align}
W = W_{\text{MSSM}} + \epsilon_i {\widehat L}_i {\widehat H}_u,
\label{eq:Wpot}
\end{align}
where $W_{MSSM}$ is the superpotential of the minimal supersymmetric
standard model (MSSM) and the parameters $\epsilon_{i}$ characterize
the bilinear R-parity violation, with the flavour index $i=1,2,3$
running over the generations.  The soft supersymmetry breaking
Lagrangian contains, in addition to the R-parity conserving operators
$V^{MSSM}_{\rm soft},$ a term associated with the R-parity violation
contribution:
\begin{align}
V_{\text{ soft}} = V^{\text{MSSM}}_{\text{soft}} + B_i \epsilon_i {\tilde L}_i H_u.
\label{eq:softWpot}
\end{align}
The $\epsilon_i$ and $B_i$ terms induce vacuum expectation values
$v_i$ for the scalar neutrinos, generating a mass mixing between
neutrinos and neutralinos. As a result, one finds that neutrino
acquires a mass at tree-level given by \cite{Hirsch:2000ef}
\begin{align}
\label{eq:mnu}
m^{\text{tree}}_\nu\approx&\frac{M_1g^2+M_2g'^2}{4\Delta_0}|\vec{\Lambda}|^2,
\end{align}
where $\Delta_0=\mbox{det}(M_{\chi_0})=-M_1M_2\mu^2+\frac{1}{2}\mu
v_dv_u(M_1g^2+M_2g'^2)$ is the determinant of the MSSM neutralino mass
matrix and $\Lambda_i=\mu v_i+v_d\epsilon_i$ is the alignment vector.
It is worth mentioning that the value of $|\vec{\Lambda}|^2$ is almost
fixed by the very precise determination of the atmospheric scale,
mainly by MINOS and other accelerator experiments.  The other two
neutrinos acquire a calculable mass only at the one-loop
level~\cite{Hirsch:2000ef,diaz:2003as} as required to explain solar
and reactor neutrino data.  The BRpV model provides a scenario where
all current neutrino oscillation data can be accounted for, i.e. it
can accommodate the required values of the neutrino mass squared
differences and mixing angles inferred from neutrino oscillation
studies~\cite{Schwetz:2011qt}.  We refer the reader to
\cite{Hirsch:2000ef,diaz:2003as} for more details about the model.

For the rest of the paper, we assume the Constrained MSSM scenario
(CMSSM) \cite{Kane:1993td} where the soft supersymmetry breaking
parameters $m_0,m_{1/2}$ and $A_0$ are assumed to be universal at the
supersymmetric Grand Unification (GUT) scale.  Thus, the model depends
upon the following eleven free parameters:
\begin{align}
m_0,\,m_{1/2},\,\tan\beta,\,\rm{sign}(\mu),\,A_0,\,\epsilon_i,\,\Lambda_i.
\end{align}
Here, $m_{1/2}$ and $m_0$ are the common gaugino mass and scalar soft
SUSY breaking masses at the unification scale, $\tan\beta$ is the
ratio between the Higgs field vacuum expectation values and $A_0$ is
the common trilinear term.

When supersymmetry is promoted to be a local symmetry of nature, the
resulting theory requires a supermultiplet which includes the
gravitino.  After supersymmetry breaking, the gravitino becomes
massive via the superhiggs mechanism.  Depending on details of the
underlying supersymmetry breaking mechanism the gravitino mass
$m_{\tilde G}$ can lie anywhere between $\mathcal{O}(\mbox{eV})$ and
$\mathcal{O}(\mbox{TeV})$. We take $m_{\tilde G}$ as a free parameter
which does not fix the scale of soft-supersymmetry breaking
parameters~\cite{Ellis:2003dn}.

\section{Gravitino cosmology }
\label{sec:gravitino}

There have been several studies of gravitino dark matter in R-parity
conserving
supersymmetry~\cite{Ellis:2003dn,Roszkowski:2004jd,Cerdeno:2005eu,Pradler:2007is}. In
this case the lightest supersymmetric particle is stable and is
potentially a viable dark matter
candidate~\cite{Ellis:2003dn,Roszkowski:2004jd,Cerdeno:2005eu,Pradler:2007is}.
Here we consider this issue within the simplest R-parity violating
scenario. 
In order to study the region of parameter space of the model which can
accomodate neutrino oscillation data, we have performed a numerical
analysis using the SPheno package~\cite{Porod:2003um}, which
calculates the renormalization group equations at two loops, generating the full supersymmetric
particle spectrum. It also includes the one-loop calculations of the
neutrino masses in the BRpV model, required in order to account for
solar neutrino conversion.  For a fixed set of CMSSM parameters
($m_0,\,m_{1/2},\,\tan\beta,\,\rm{sign}(\mu),\,A_0$) we determine the
set of R-parity breaking parameters $\epsilon_i$ and $\Lambda_i$
responsible for generating neutrino mass squared differences and
mixing angles consistent at 3$\sigma$ with the measured values
\cite{Schwetz:2011qt}.  We repeat this procedure fixing $A_0=-100$ GeV
sign$(\mu)=+1$ and scanning over the other CMSSM parameters in the
range:
\begin{eqnarray}
\label{sec:grav-dark-matt}
&240 \leq m_{1/2}\leq 3000 \mbox{ GeV}, & \\ \nonumber
&3\leq\tan\beta\leq50, &\\ 
&200\leq M_0\leq 1000 \mbox{ GeV}.& \nonumber
\end{eqnarray}

\subsection{Gravitino dark matter relic density}
\label{sec:relic-density}

Gravitinos are produced in the early Universe after the reheating
phase through particles scattering occurring in the thermal plasma,
the dominant contribution coming from SUSY quantum chromodinamycs
processes
\cite{Bolz:2000fu,Pradler:2006qh,Pradler:2006hh,Khlopov:1984pf,Rychkov:2007uq}.  The
gravitino relic abundance critically depends on the reheating
temperature $T_R$ and it is given by:
\cite{Bolz:2000fu,Pradler:2006qh} \begin{align} \Omega_{\tilde G}h^2=
  &\sum_{i=1}^3\omega_i
  \left[g_i(T_R)\right]^2\left(1+\frac{\left[M_i(T_R)\right]^2}{3m_{\tilde
        G}^2}\right) {\rm ln}\left(\frac{k_i}{g_i(T_R)}\right)
  \left(\frac{m_{\tilde G}}{100\mbox{
        GeV}}\right)\left(\frac{T_R}{10^{10}\mbox{ GeV}}\right),
\label{eq:gdensity}
\end{align}
with $M_i(T_R)$ and $g_i(T_R)$ respectively the gaugino mass
parameters and gauge coupling constants at $T_R$ energy scale.  The
index $i$ runs over the Standard Model gauge group factors and the
constants $\omega_i$ and $k_i$ are $\omega_i=(0.018,0.044,0.117)$ and
$k_i=(1.266, 1.312, 1.271)$.  The gaugino masses and the gauge
coupling constants can be evaluated at the energy scale $T_R$ using
the renormalization group equations (RGEs), which at one-loop level
are given as
\begin{align}
 g_i(T_R)=&\left[g_i(m_Z)^{-2}-\frac{\beta_i^{(1)}}{8\pi^2}{\rm ln}\left(\frac{T_R}{m_Z}\right)\right]^{-1/2},\\
 M_i(T_R)=&\left(\frac{g_i(T_R)}{g_i(m_Z)}\right)^2M_i(m_Z).
\end{align}
In the MSSM, the beta function coefficients are
$\beta_i^{(1)}=(11,1,-3)$.  Assuming universal gaugino soft masses at
the supersymmetric GUT scale, their values at the electroweak scale
($m_Z$) follow the relations $M_3\simeq 3.1M_2 \simeq 5.9 M_1.$

\begin{figure}[ht]
\begin{center}
 \includegraphics[scale=1.0]{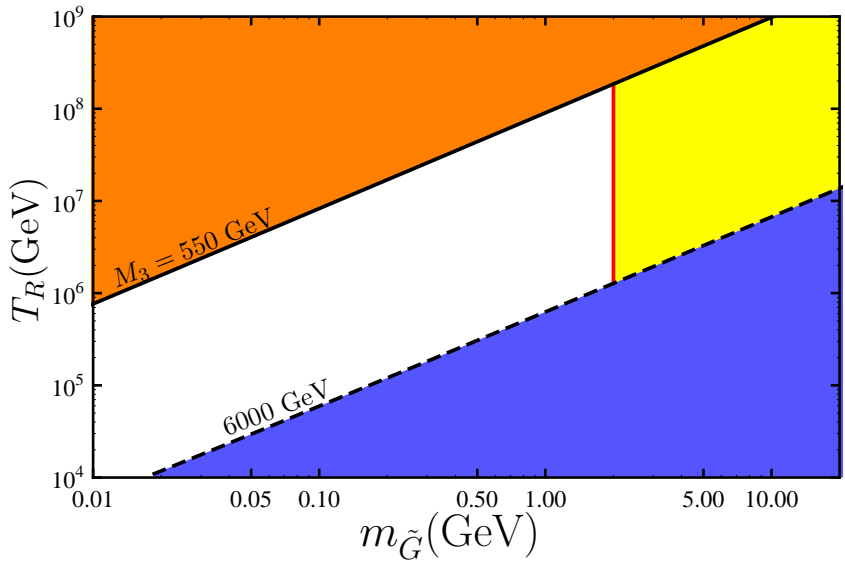}
 \caption{Region (white) in the $(m_{\tilde G}, T_R)$ plane consistent
   with neutrino oscillations, gravitino dark matter, gamma-ray line
   searches, and gluino searches.  In the orange area a consistent
   gravitino relic abundance would require gluino masses excluded by
   present collider searches. In contrast, the blue area is viable but
   requires too large gluino masses, $M_3>6000$ GeV. The yellow region
   is excluded by astrophysical gamma-ray line searches.}
\label{fig:tr-mg}
\end{center}
\end{figure}
In figure~\ref{fig:tr-mg} the black lines show the combination of
gravitino masses and reheating temperatures for which the gravitino
relic abundance is consistent with the dark matter density inferred
from astrophysical observations \cite{Komatsu:2010fb},
$\Omega_{DM}h^2=0.1123$. There, we have assumed gaugino universality
and the two curves correspond to two different values for the soft
gluino mass parameter $M_3$.  
The orange region (upper dark gray) requires values
of $M_3$ smaller than $550$ GeV in order to obtain the correct
gravitino relic abundance consistent with WMAP. Present bounds on the
gluino mass from recent searches at the LHC already exclude this
region~\cite{Chatrchyan:2011wc,Khachatryan:2011tk,Chatrchyan:2011bj,Chatrchyan:2011ah,daCosta:2011qk,Aad:2011ks,Aad:2011xm}.
On the other hand, the blue area (lower dark grey) is viable but
corresponds to $M_{3}\gtrsim 6000$~GeV which, though
phenomenologically acceptable, is theoretically disfavored if
supersymmetry is supposed to ``protect'' the hierarchical problem.  

We notice that for gravitinos in the mass range up to few GeV, the
corresponding values of reheating temperature are relatively low and
compatible with the lower bounds that can be inferred from cosmic
microwave background radiation observations~\cite{Martin:2010kz}. 

\subsection{Gravitino decays}
\label{sec:gravitino-decays-1}

In the presence of R-parity breaking, as in the bilinear R-parity
model we have considered above, the LSP decays.  In particular, if the
strength of the bilinear R-parity violating parameters is chosen so as
to reproduce the neutrinos masses and mixing angles indicated by
neutrino oscillation experiments~\cite{Schwetz:2011qt}, a neutralino
LSP would decay with a lifetime way too short when compared with the
age of the Universe.  However, if the LSP is a gravitino, the double
suppression provided by the smallness of the R parity violating
parameters and the Planck-scale suppression of the coupling governing
the decay rate greatly increase its lifetime, making it a perfect dark
matter candidate~\cite{Takayama:2000uz,Buchmuller:2007ui}. It would
also provide an example of the generic expectation that gravitational
interactions break global
symmetries~\cite{coleman:1988tj,Holman:1992us}, in this case R-parity
and lepton number.

The unstable gravitino dark matter scenarios can potentially be tested
with indirect astrophysical dark matter searches.  In the bilinear
R-parity breaking model under consideration, the gravitino decays as
${\tilde G}\to \nu \gamma$ with a width:
\begin{equation}
\Gamma=\Gamma({\tilde G}\to \sum_i \nu_i \gamma) \simeq \frac{1}{32 \pi}
|U_{\tilde{\gamma} \nu}|^2 \frac{m_{\tilde G}^3}{M_{P}^2},
\label{eq:GravDec}
\end{equation}
where the R-parity breaking mixing parameter $|U_{\tilde{\gamma}
  \nu}|^2$ from the $7\times 7$ neutralino mixing matrix, is
\cite{Hirsch:2005ag}
\begin{align}
  |U_{{\tilde \gamma} \nu}|^2=\sum_{a=i+4}|\cos\theta_W N_{a1}+\sin\theta_W N_{a2}|^2\,,
\end{align}
where the $N$-coefficients denote the neutrino projections onto the
gauginos.  Following \cite{diaz:2003as} this can be calculated
perturbatively as
 \begin{align}
\label{eq:U}
  |U_{{\tilde \gamma} \nu}|^2\approx\frac{\mu^2 g^2 \sin^2\theta_W}{4\Delta_0^2}(M_2-M_1)^2|\vec{\Lambda}|^2,
 \end{align}
 with $\Delta_0=\mbox{det}(M_{\chi_0})=-M_1M_2\mu^2+\frac{1}{2}\mu v_d
 v_u(M_1g^2+M_2g'^2)$ and $\Lambda_i=\mu v_i+v_d\epsilon_i$.

 For each set of parameters generated through the scanning procedure
 given in Eq.~(\ref{sec:grav-dark-matt}) yielding the correct values
 of the neutrino oscillation parameters, we compute the gravitino
 lifetime, using equations (\ref{eq:GravDec}) and (\ref{eq:U}).
 This decay mode is particularly interesting from the point of view of
 indirect dark matter detection.  Indeed monochromatic photons of
 $\sim \mbox{GeV}$ energies are generally not expected to be produced
 by conventional astrophysical processes.  For this reason, the
 detection of gamma-ray lines would be a striking signature of dark
 matter processes, pointing either to annihilations
 \cite{Bergstrom:1997fh,Gustafsson:2007pc,Bertone:2009cb,Jackson:2009kg,Bertone:2010fn,Dudas:2009uq,Mambrini:2009ad}
 or to decays of dark matter particles
 \cite{Ibarra:2007wg,Arina:2009uq,Garny:2010eg,Bobrovskyi:2010ps,Choi:2009ng,Bajc:2010qj}.
 Search of gamma-ray lines have been recently performed using the data
 of the Fermi-LAT satellite~\cite{Abdo:2010nc,Vertongen:2011mu}.  The
 derived upper bounds on the gamma-ray line fluxes can be used to
 constrain the unstable gravitino dark matter model under
 consideration. 

\begin{figure}[ht]
\begin{center}
 \includegraphics[scale=0.99]{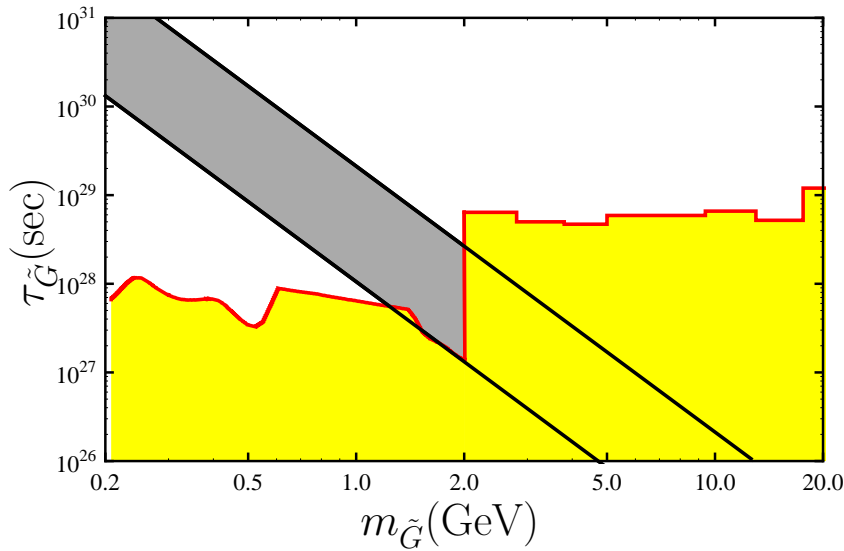}
 \caption{Allowed gravitino mass-lifetime region (grey color)
   consistent with neutrino oscillation data and astrophysical bounds
   on gamma-ray lines from dark matter decay. The yellow region is
   excluded by gamma-ray line searches (Fermi and EGRET constraints
   are respectively above and below $1$ GeV).  The lower and upper black lines 
   correspond to $m_{1/2}=240$  and $3000$ GeV respectively.} 
\label{fig:ltg-mg}
\end{center}
\end{figure}

 In figure~\ref{fig:ltg-mg} we present the lower bounds on the
 gravitino lifetime for dark matter gravitinos decaying into $\nu
 \gamma$~\cite{Vertongen:2011mu}. These constraints have been computed
 assuming a Navarro-Frenk-White (NFW) dark matter density
 profile~\cite{Navarro:1996gj} and for the region of observation
 dubbed as "Halo" in Ref.~\cite{Vertongen:2011mu}.  The bounds are not
 too sensitive to the exact shape of the dark matter profiles or
 region of observation considered.  At energies below $\sim 1$ GeV we
 consider the bounds on gamma-ray lines obtained in
 Ref.~\cite{Pullen:2006sy} by analyzing the data from EGRET.  We
 traslate the upper limits on the gamma-ray line fluxes into bounds on
 the gravitino lifetime for gravitino decays into $\nu
 \gamma$~\cite{Pullen:2006sy}.  We consider a NFW density distribution
 while a shallower isothermal profile would lead to a bound a factor
 two less stringent.
Finally, we note that gravitino decays into three body final states
could be relevant for gravitino masses larger than those required in
our case~\cite{Choi:2010jt,Diaz:2011pc}.

  The area between the two black lines in figure~\ref{fig:ltg-mg}
  corresponds to the region of the parameters compatible with neutrino
  physics.  We notice that the two curves correspond approximately to
  constant values of $m_{1/2}=240$ GeV (lower line) and
  $m_{1/2}=3000$~GeV (upper line).  Indeed, once the constraints from
  neutrino oscillations are imposed, the matrix element
  $|U_{\tilde{\gamma} \nu}|^2$ determining the gravitino lifetime,
  depends mostly on $m_{1/2}.$ Universal gaugino masses of
  $m_{1/2}=240$ GeV and $m_{1/2}=3000$~GeV lead at the scale $M_Z$ to
  $M_3\approx~550$~GeV and $M_3\approx6000$~GeV respectively.

  Taking into account the Fermi-LAT and EGRET bounds on gamma-ray
  lines from dark matter decay (yellow region in
  figure~\ref{fig:ltg-mg}) and assuming the gravitino as dark matter
  particle, we can derive an upper bound on the gravitino mass of the
  order $m_{\tilde G}\sim 2$ GeV.  Thus, the grey area in
  figure~\ref{fig:ltg-mg} corresponds to the region of the parameter
  space which simultaneusly explains neutrino oscillation data and
  satisfies the constraints from gamma-ray line searches.

We now translate the bounds from neutrino oscillation physics and
gamma-ray line searches to the $(m_{\tilde G},T_R$) plane shown in
figure~\ref{fig:tr-mg}. They correspond to the yellow area (light
grey). 
We see that assuming the gravitino as dark matter candidate in BRpV
and imposing the constraints from neutrino oscillations and gamma-ray
line searches we can derive an upper bound on reheating temperature of
the order $T_R\sim 10^8$ GeV and an upper bound on the gravitino mass
of the order $m_{\tilde G}\sim 2$ GeV.
Similar results are expected in other R-parity violation schemes, such
as considered in Ref.~\cite{Choi:2009ng}.

\subsection{Non-universal gaugino masses}
\label{noU}
\begin{figure}[ht] 
\begin{center}
 \includegraphics[scale=0.5]{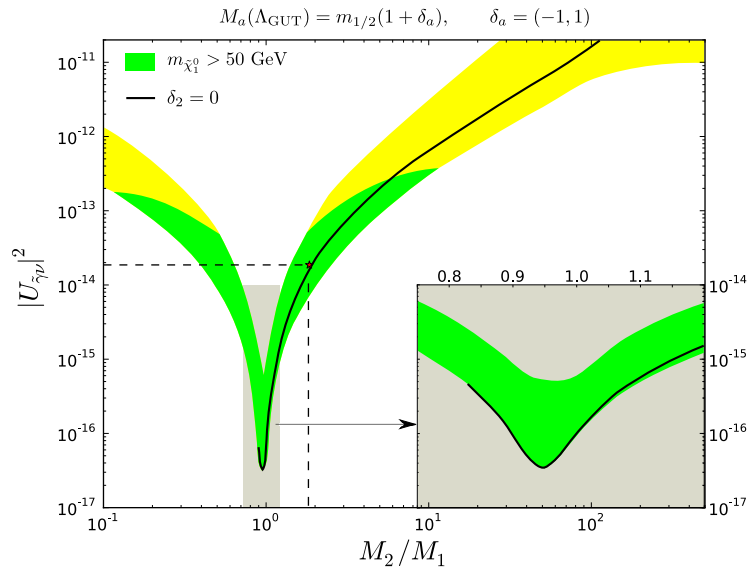}
 \caption{Neutrino-photino mixing $|U_{{\tilde \gamma}\nu}|^2$ as a
   function of the low energy ratio $M_2/M_1$ when
   $M_a=m_{1/2}(1+\delta_a)$ have been assumed at the unification
   scale (with $\delta_3=0)$). We have set $m_{1/2}=500\ $GeV,
   $m_0=1000\ $GeV, $A_0=-100\ $GeV, $\tan\beta=10$ and
   $\operatorname{sign}(\mu)=+1$. In the green (dark gray) region
   $m_{\tilde{\chi}_1^0}>50\ $GeV. The explicit minimum of
   $|U_{{\tilde \gamma}\nu}|^2$ is shown in the zoomed gray area.}
\label{fig:U2}
\end{center}
\end{figure}

We now study the effects of non-universal gaugino masses on the
gravitino mass upper bound. The non-universality effects enter in the
gravitino lifetime through the neutrino-``photino'' mixing parameter
$|U_{{\tilde \gamma}\nu}|^2$. 
At the unification scale non-universal gaugino masses can be
parametrized as \cite{Akula:2011dd}
\begin{align}
M_a=m_{1/2}(1+\delta_a),
\end{align}
with the parameters $\delta_a$, $a=1,2,3$, characterizing the
deviation from universality. For illustratation we choose the ranges
$\delta_{1,2}=(-1,1)$, keeping $\delta_3=0$, and fix a typical CMSSM
point satisfying all the phenomenological constraints with
$m_{1/2}=500$~GeV, $m_0=1000$~GeV, $A_0=-100$, $\tan\beta=10$, and
$\operatorname{sgn}(\mu)>0$. Then we use SPheno with a random set of
$\delta_a$ values to calculate $|U_{\tilde{\gamma}\nu}|^2$, with the
best possible fit to neutrino masses and mixings for each point of the
scan. We also check that limits on sparticle searches are obeyed,
e.~g. $m_{\tilde{\chi}^\pm_1}>103$\ GeV, and that a non-bino
neutralino has a mass larger than $50$~GeV
\cite{Akula:2011dd,AristizabalSierra:2008ye}. The results are shown in
the yellow (light gray) region of figure~\ref{fig:U2} where the ratio
$M_2/M_1$ is calculated at the electroweak scale. In the green (dark
gray) region the mass neutralino is larger than $50$~GeV, while in the
solid black line $\delta_2=0$. 

From the approximate expression for $|U_{\tilde{\gamma}\nu}|^2$ in
eq.~\eqref{eq:U} one would expect vanishing values at $M_2\approx
M_1$. However, from the loop-corrected neutralino mass matrix
calculated from SPheno, one obtains that $|U_{\tilde{\gamma}\nu}|^2$
has a minimum non-zero value at $M_2\approx M_1$. In figure
\ref{fig:U2}, the dashed lines indicate the point where the gaugino
masses arises from universal conditions. From the gravitino lifetime
in \eqref{eq:GravDec}, one sees that for neutralino masses larger
(smaller) than the CMSSM reference value, the gravitino mass bound is
weaker (stronger). For example, for $m_{1/2}=500$~GeV, we have
$|U_{{\tilde \gamma}\nu}|^2\approx1.6\times10^{-14}$. From
eq. (\ref{eq:GravDec}) the maximum gravitino mass can be expressed as
\begin{align}
\label{eq:bound}
  m^{\text{max}}_{\tilde G}\approx  2\,\mbox{GeV}\left(\frac{ 3\times
      10^{27}\mbox{s}}{\tau^{\text{min}}_{\tilde
        G}}\right)^{1/3}\left(\frac{1.6\times
      10^{-14}}{|U_{\tilde{\gamma}\nu}|^2}\right)^{1/3},
\end{align}
where $\tau^{min}_{\tilde G}$ is the minimum gravitino lifetime
allowed by gamma-ray line searches (see figure
\ref{fig:ltg-mg}). When $M_2\approx M_1$, we obtain the minimum value
for $|U_{{\tilde \gamma}\nu}|^2\approx 4\times 10^{-17}$, and an upper
bound on gravitino mass of order $m_{\tilde G}\sim 7$ GeV. This
illustrates the relative robustness of the gravitino mass bound
against deviations from gaugino universality.

While this holds for a given CMSSM point, we have not been able to
find other points in parameter space where the gravitino mass bound
changes by more than an order of magnitude. The point is that, even
though radiative effects in the neutralino mass matrix may change by
three orders of magnitude (see figure~\ref{fig:U2}) the bound changes
only as the 1/3 power of that, according to eq.~(\ref{eq:bound}),
hence is relative stable.

Let's now briefly comment about implications on Big Bang
Nucleosynthesis (BBN).  In R-parity conserving scenarios with
gravitino dark matter, the neutralino has a large lifetime since its
decays into the LSP are suppressed by the Planck scale thus it may
decay during the Big Bang Nucleosynthesys epoch, spoiling its
predictions \cite{Kawasaki:2004qu} (BBN demands a NLSP lifetime less
than 0.1 s \cite{Kawasaki:2004qu}).
In contrast, in the model under consideration, the next to lightest supersymmetric particle (NLSP) 
decays occur well before the BBN epoch because of the presence of the gravity
unsuppressed R-parity violating interactions, keeping therefore the
successful BBN predictions of the light element abundances.

\section{Prospects for collider seaches}
\label{sec:nnlsp}
When R-parity is conserved, all supersymmetric particles undergo
cascade decays to the  next to lightest supersymmetric particle,
which subsequently decays (of course, with gravitational strength) to
the gravitino.
The implications for collider searches and cosmology strongly depend
on which superpartner is the NLSP. In BRpV models, in addition to
generating the neutrino masses, the neutralino-neutrino mixing also
induces NLSP decays into Standard Model particles, strongly correlated
with the neutrino oscillation
parameters~\cite{Porod:2000hv,Choi:1999tq,Mukhopadhyaya:1998xj}. Since
R-parity violating couplings are not so small, displaced vertices are
expected in the NLSP decay
\cite{deCampos:2005ri,deCampos:2007bn,DeCampos:2010yu}. In what
follows we will consider the neutralino as the NLSP.

\begin{figure}[ht] 
\begin{center}
 \includegraphics[scale=0.8]{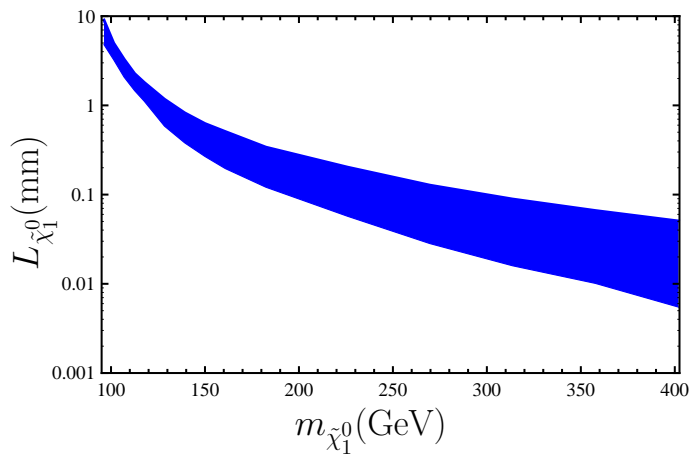}
 \caption{Neutralino NLSP decay length as a function of its mass. For
   illustration we show the results of a scan with $A_0=-100$ GeV,
   sign$(\mu)=+1$, $\tan\beta=10$, $200\leq M_0\leq 1000 \mbox{ GeV}$
   and $240 \leq m_{1/2}\leq 1000 \mbox{ GeV}$.}
\label{fig:declen}
\end{center}
\end{figure}

In R-parity conserving scenarios the neutralino as the NLSP has the
following decay channels \cite{Giudice:1998bp}
\begin{align}\label{eq:chRpC}\nonumber
& \tilde{\chi}_1^0\to \gamma{\tilde G},\nonumber\\
& \tilde{\chi}_1^0\to Z{\tilde G},\nonumber\\
& \tilde{\chi}_1^0\to h^0{\tilde G},
\end{align}
In the presence of R-parity breaking additional decay channels
exist~\cite{Porod:2000hv}, namely,
\begin{align}\label{eq:chRpV}\nonumber
& \tilde{\chi}_1^0\to h^{0}\nu_i,\nonumber\\
& \tilde{\chi}_1^0\to \gamma\nu_i,\nonumber\\
& \tilde{\chi}_1^0\to W^{\pm}l_i^{\mp},\nonumber\\
& \tilde{\chi}_1^0\to Z^{0}\nu_i.
\end{align} 
Neutralino can also decay to three fermions by scalar quark and scalar
lepton exchange in R-parity violating models. The three channels in
(\ref{eq:chRpC}) are Planck-mass-suppressed and, for the gravitino
mass range of interest, are negligible compared with those in
eq.~(\ref{eq:chRpV}). Indeed, $Br(\tilde{\chi}_1^0\to \gamma{\tilde
  G})<10^{-5}$ for $m_{\tilde G}> 10$ keV \cite{Hirsch:2005ag}. The
decay into the Higgs boson is scalar mixing suppressed, while the
radiative channel is loop suppressed.
As a result, decays to gauge bosons are dominant for large $m_0$.
Therefore these collider signals are independent of the fact that the
gravitino is lightest supersymmetric particle or not.  In particular,
the predictions at colliders for a neutralino LSP in the CMSSM with
BRpV studied in
\cite{Porod:2000hv,Choi:1999tq,Mukhopadhyaya:1998xj,deCampos:2005ri,deCampos:2007bn,DeCampos:2010yu},
such as the displaced vertex signals illustrated in figure
\ref{fig:declen}, remain unchanged.

\section{Summary}
\label{sec:Summary}
We have considered the CMSSM model with bilinear R-parity violation
with gravitino as LSP.
By imposing the constraints from neutrino oscillation data, dark
matter relic density and gamma-ray line searches at Fermi-LAT and
EGRET we have shown that the gravitino does provide a viable
radiatively decaying dark matter particle, provided its mass and
reheat temperature are bounded as $m_{\tilde G} \lsim$ 1-10~GeV and
$T_R \lsim$ few $\times 10^7-10^8$~GeV (as we saw the bounds get
looser if the universality hypothesis in gaugino masses is relaxed).
The expected signatures associated to the NLSP decays at collider
experiments like the Large Hadron Collider do not depend on the
presence of the gravitino and so are the same as those previously
studied~\cite{Porod:2000hv,Choi:1999tq,Mukhopadhyaya:1998xj,deCampos:2005ri,deCampos:2007bn,DeCampos:2010yu}. In
particular neutrino mass and mixing parameters may be reconstructed at
accelerator experiments by measuring the ratio of semileptonic
neutralino decays branching ratios induced by the charged current.

\section*{Acknowledgments}
We acknowledge German Gomez-Vargas for useful discussions.
This work was supported by the Spanish MICINN under grants
FPA2008-00319/FPA and MULTIDARK Consolider CSD2009-00064, by
Prometeo/2009/091, by the EU grant UNILHC PITN-GA-2009-237920 and by
the EIA grant CI-2009-2. O. Z. acknowledges to AHEP group and IFIC for
their hospitality and support. D.R was partly 
supported by Sostenibilidad-UdeA/2009 grant: IN10140-CE
\bibliographystyle{h-physrev5}

\begin{thebibliography}{10}

\bibitem{berezinsky:1993fm}
V.~Berezinsky and J.~W.~F. Valle,
\newblock Phys. Lett. {\bf B318}, 360 (1993), hep-ph/9309214.

\bibitem{lattanzi:2007ux}
M.~Lattanzi and J.~W.~F. Valle,
\newblock Phys. Rev. Lett. {\bf 99}, 121301 (2007), arXiv:0705.2406 [astro-ph].

\bibitem{bazzocchi:2008fh}
F.~Bazzocchi {\em et~al.},
\newblock JCAP {\bf 0808}, 013 (2008), arXiv:0805.2372.

\bibitem{Esteves:2010sh}
J.~Esteves {\em et~al.},
\newblock Phys.Rev. {\bf D82}, 073008 (2010), arXiv:1007.0898.

\bibitem{hirsch:2004he}
M.~Hirsch and J.~W.~F. Valle,
\newblock New J. Phys. {\bf 6}, 76 (2004), hep-ph/0405015.

\bibitem{masiero:1990uj}
A.~Masiero and J.~W.~F. Valle,
\newblock Phys. Lett. {\bf B251}, 273 (1990).

\bibitem{Diaz:1997xc}
M.~A. Diaz, J.~C. Romao, and J.~W.~F. Valle,
\newblock Nucl. Phys. {\bf B524}, 23 (1998).

\bibitem{Hirsch:2000ef}
M.~Hirsch {\em et~al.},
\newblock Phys. Rev. {\bf D62}, 113008 (2000), hep-ph/0004115,
\newblock Err-ibid. {\bf D65}:119901,2002.

\bibitem{diaz:2003as}
M.~A. Diaz {\em et~al.},
\newblock Phys. Rev. {\bf D68}, 013009 (2003), hep-ph/0302021.

\bibitem{Takayama:2000uz}
F.~Takayama and M.~Yamaguchi,
\newblock Phys. Lett. {\bf B485}, 388 (2000), arXiv:hep-ph/0005214.

\bibitem{Buchmuller:2007ui}
W.~Buchmuller, L.~Covi, K.~Hamaguchi, A.~Ibarra, and T.~Yanagida,
\newblock JHEP {\bf 0703}, 037 (2007), arXiv:hep-ph/0702184.

\bibitem{Hirsch:2005ag}
M.~Hirsch, W.~Porod, and D.~Restrepo,
\newblock JHEP {\bf 03}, 062 (2005), hep-ph/0503059.

\bibitem{Schwetz:2011qt}
T.~Schwetz, M.~Tortola, and J.~W.~F. Valle,
\newblock New J. Phys. {\bf 13}, 063004 (2011), and T2K/MINOS update 
in addendum in New J.Phys. {\bf 13}, 109401 (2011);
for references to other groups see
New J.\ Phys.\  {\bf 10}, 113011 (2008),
and M.~Maltoni et al, New J.\ Phys.\  {\bf 6}, 122 (2004).

\bibitem{Abdo:2010nc}
A.~Abdo {\em et~al.},
\newblock Phys.Rev.Lett. {\bf 104}, 091302 (2010), arXiv:1001.4836.

\bibitem{Vertongen:2011mu}
G.~Vertongen and C.~Weniger,
\newblock JCAP {\bf 1105}, 027 (2011), arXiv:1101.2610,

\bibitem{Mukhopadhyaya:1998xj}
B.~Mukhopadhyaya, S.~Roy, and F.~Vissani,
\newblock Phys. Lett. {\bf B443}, 191 (1998).

\bibitem{Choi:1999tq}
S.~Y. Choi, E.~J. Chun, S.~K. Kang, and J.~S. Lee,
\newblock Phys. Rev. {\bf D60}, 075002 (1999), hep-ph/9903465.

\bibitem{Porod:2000hv}
W.~Porod {\em et~al.},
\newblock Phys. Rev. {\bf D63}, 115004 (2001).

\bibitem{deCampos:2005ri}
F.~de~Campos {\em et~al.},
\newblock Phys. Rev. {\bf D71}, 075001 (2005), hep-ph/0501153.

\bibitem{deCampos:2007bn}
F.~de~Campos {\em et~al.},
\newblock JHEP {\bf 05}, 048 (2008).

\bibitem{DeCampos:2010yu}
F.~De~Campos {\em et~al.},
\newblock Phys. Rev. {\bf D82}, 075002 (2010), arXiv:1006.5075.

\bibitem{Kane:1993td}
G.~L. Kane, C.~F. Kolda, L.~Roszkowski, and J.~D. Wells,
\newblock Phys.Rev. {\bf D49}, 6173 (1994), arXiv:hep-ph/9312272.

\bibitem{Ellis:2003dn}
J.~R. Ellis, K.~A. Olive, Y.~Santoso, and V.~C. Spanos,
\newblock Phys.Lett. {\bf B588}, 7 (2004), arXiv:hep-ph/0312262.

\bibitem{Roszkowski:2004jd}
L.~Roszkowski, R.~Ruiz~de Austri, and K.-Y. Choi,
\newblock JHEP {\bf 0508}, 080 (2005), arXiv:hep-ph/0408227.

\bibitem{Cerdeno:2005eu}
D.~G. Cerdeno, K.-Y. Choi, K.~Jedamzik, L.~Roszkowski, and R.~Ruiz~de Austri,
\newblock JCAP {\bf 0606}, 005 (2006), arXiv:hep-ph/0509275.

\bibitem{Pradler:2007is}
J.~Pradler and F.~D. Steffen,
\newblock Phys.Lett. {\bf B666}, 181 (2008), arXiv:0710.2213.

\bibitem{Porod:2003um}
W.~Porod,
\newblock Comput. Phys. Commun. {\bf 153}, 275 (2003), hep-ph/0301101.

\bibitem{Bolz:2000fu}
M.~Bolz, A.~Brandenburg, and W.~Buchmuller,
\newblock Nucl.Phys. {\bf B606}, 518 (2001), arXiv:hep-ph/0012052.

\bibitem{Pradler:2006qh}
J.~Pradler and F.~D. Steffen,
\newblock Phys.Rev. {\bf D75}, 023509 (2007), arXiv:hep-ph/0608344.

\bibitem{Pradler:2006hh}
J.~Pradler and F.~D. Steffen,
\newblock Phys.Lett. {\bf B648}, 224 (2007), arXiv:hep-ph/0612291.


\bibitem{Khlopov:1984pf}
  M.~Y.~.Khlopov and A.~D.~Linde,
\newblock Phys.\ Lett.\ B {\bf 138}, 265 (1984).

\bibitem{Rychkov:2007uq}
V.~S. Rychkov and A.~Strumia,
\newblock Phys.Rev. {\bf D75}, 075011 (2007), arXiv:hep-ph/0701104.

\bibitem{Komatsu:2010fb}
WMAP collaboration, E.~Komatsu {\em et~al.},
\newblock Astrophys.J.Suppl. {\bf 192}, 18 (2011), arXiv:1001.4538.

\bibitem{Chatrchyan:2011wc}
CMS Collaboration, S.~Chatrchyan {\em et~al.},
\newblock Phys. Rev. Lett. {\bf 106}, 211802 (2011), arXiv:1103.0953.

\bibitem{Khachatryan:2011tk}
CMS collaboration, V.~Khachatryan {\em et~al.},
\newblock Phys. Lett. {\bf B698}, 196 (2011), arXiv:1101.1628.

\bibitem{Chatrchyan:2011bj}
CMS collaboration, S.~Chatrchyan {\em et~al.},
\newblock JHEP {\bf 07}, 113 (2011), arXiv:1106.3272.

\bibitem{Chatrchyan:2011ah}
CMS collaboration, S.~Chatrchyan {\em et~al.},
\newblock JHEP {\bf 06}, 093 (2011), arXiv:1105.3152.

\bibitem{daCosta:2011qk}
ATLAS collaboration, J.~B.~G. da~Costa {\em et~al.},
\newblock Phys. Lett. {\bf B701}, 186 (2011), arXiv:1102.5290.

\bibitem{Aad:2011ks}
ATLAS collaboration, G.~Aad {\em et~al.},
\newblock Phys. Lett. {\bf B701}, 398 (2011), arXiv:1103.4344.

\bibitem{Aad:2011xm}
ATLAS collaboration, G.~Aad {\em et~al.},
\newblock Eur. Phys. J. {\bf C71}, 1682 (2011), arXiv:1103.6214.

\bibitem{Martin:2010kz}
J.~Martin and C.~Ringeval,
\newblock Phys.Rev. {\bf D82}, 023511 (2010), arXiv:1004.5525.

\bibitem{coleman:1988tj}
S.~R. Coleman,
\newblock Nucl. Phys. {\bf B310}, 643 (1988).

\bibitem{Holman:1992us}
R.~Holman {\em et~al.},
\newblock Phys.Lett. {\bf B282}, 132 (1992), arXiv:hep-ph/9203206.

\bibitem{Bergstrom:1997fh}
L.~Bergstrom and P.~Ullio,
\newblock Nucl.Phys. {\bf B504}, 27 (1997), arXiv:hep-ph/9706232.

\bibitem{Gustafsson:2007pc}
M.~Gustafsson, E.~Lundstrom, L.~Bergstrom, and J.~Edsjo,
\newblock Phys.Rev.Lett. {\bf 99}, 041301 (2007), arXiv:astro-ph/0703512.

\bibitem{Bertone:2009cb}
G.~Bertone, C.~Jackson, G.~Shaughnessy, T.~M. Tait, and A.~Vallinotto,
\newblock Phys.Rev. {\bf D80}, 023512 (2009), arXiv:0904.1442.

\bibitem{Jackson:2009kg}
C.~Jackson, G.~Servant, G.~Shaughnessy, T.~M. Tait, and M.~Taoso,
\newblock JCAP {\bf 1004}, 004 (2010), arXiv:0912.0004.

\bibitem{Bertone:2010fn}
G.~Bertone, C.~Jackson, G.~Shaughnessy, T.~M. Tait, and A.~Vallinotto,
\newblock (2010), arXiv:1009.5107.

\bibitem{Dudas:2009uq}
E.~Dudas, Y.~Mambrini, S.~Pokorski, and A.~Romagnoni,
\newblock JHEP {\bf 0908}, 014 (2009), arXiv:0904.1745.

\bibitem{Mambrini:2009ad}
Y.~Mambrini,
\newblock JCAP {\bf 0912}, 005 (2009), arXiv:0907.2918.

\bibitem{Ibarra:2007wg}
A.~Ibarra and D.~Tran,
\newblock Phys.Rev.Lett. {\bf 100}, 061301 (2008), arXiv:0709.4593.

\bibitem{Arina:2009uq}
C.~Arina, T.~Hambye, A.~Ibarra, and C.~Weniger,
\newblock JCAP {\bf 1003}, 024 (2010), arXiv:0912.4496.

\bibitem{Garny:2010eg}
M.~Garny, A.~Ibarra, D.~Tran, and C.~Weniger,
\newblock JCAP {\bf 1101}, 032 (2011), arXiv:1011.3786.

\bibitem{Bobrovskyi:2010ps}
S.~Bobrovskyi, W.~Buchmuller, J.~Hajer, and J.~Schmidt,
\newblock JHEP {\bf 1010}, 061 (2010), arXiv:1007.5007.

\bibitem{Choi:2009ng}
K.-Y. Choi, D.~E. Lopez-Fogliani, C.~Munoz, and R.~R. de~Austri,
\newblock JCAP {\bf 1003}, 028 (2010), arXiv:0906.3681.


\bibitem{Bajc:2010qj}
B.~Bajc, T.~Enkhbat, D.~K.~Ghosh, G.~Senjanovic and Y.~Zhang,
\newblock  JHEP {\bf 1005}, 048 (2010), arXiv:1002.3631.


\bibitem{Navarro:1996gj}
J.~F. Navarro, C.~S. Frenk, and S.~D.~M. White,
\newblock Astrophys. J. {\bf 490}, 493 (1997), arXiv:astro-ph/9611107.

\bibitem{Pullen:2006sy}
A.~R. Pullen, R.-R. Chary, and M.~Kamionkowski,
\newblock Phys.Rev. {\bf D76}, 063006 (2007), arXiv:astro-ph/0610295.

\bibitem{Choi:2010jt}
K.-Y. Choi, D.~Restrepo, C.~E. Yaguna, and O.~Zapata,
\newblock JCAP {\bf 1010}, 033 (2010), arXiv:1007.1728.

\bibitem{Diaz:2011pc}
M.~A. Diaz, S.~G. Saenz, and B.~Koch,
\newblock Phys.\ Rev.\ D {\bf 84}, 055007 (2011) 



\bibitem{Akula:2011dd}
S.~Akula, D.~Feldman, Z.~Liu, P.~Nath, and G.~Peim,
\newblock Mod. Phys. Lett. {\bf A26}, 1521 (2011), arXiv:1103.5061.

\bibitem{AristizabalSierra:2008ye}
D.~Aristizabal~Sierra, W.~Porod, D.~Restrepo, and C.~E. Yaguna,
\newblock Phys. Rev. {\bf D78}, 015015 (2008), arXiv:0804.1907.

\bibitem{Kawasaki:2004qu}
M.~Kawasaki, K.~Kohri, and T.~Moroi,
\newblock Phys. Rev. {\bf D71}, 083502 (2005), astro-ph/0408426.

\bibitem{Giudice:1998bp}
G.~F. Giudice and R.~Rattazzi,
\newblock Phys. Rept. {\bf 322}, 419 (1999), hep-ph/9801271.

\end{thebibliography}

\end{document}